\documentclass[aps,pre, floatfix, twocolumn, amsmath, showpacs,superscriptaddress, 10pt]{revtex4-1} 
\usepackage[pdftex]{graphicx}            % to include images
\usepackage{epstopdf}
\usepackage{natbib}
\usepackage{amsfonts}
\usepackage{color}
\usepackage{amsmath}
\newcommand{\etal}{et al.~}  
  
\newcommand{\rvec}{\textbf{r}}  
\newcommand{\rvecp}{\textbf{r}^{\prime}}
\newcommand{\thetap}{\theta^{\prime}}   
\newcommand{\phip}{\phi^{\prime}}

\begin{document}

\title{\Large Chimera states on the surface of a sphere} 

\author{Mark J. Panaggio}
\email[email: ]{markpanaggio2014@u.northwestern.edu}
\affiliation{Department of Engineering Sciences and Applied Mathematics, Northwestern University, Evanston, Illinois 60208, USA}
\affiliation{Mathematics Department, Rose-Hulman Institute of Technology, Terre Haute, Indiana 47803, USA}
\author{Daniel M. Abrams}
\affiliation{Department of Engineering Sciences and Applied Mathematics, Northwestern University, Evanston, Illinois 60208, USA}
\affiliation{Department of Physics and Astronomy, Northwestern University, Evanston, Illinois 60208, USA}
\affiliation{Northwestern Institute on Complex Systems, Northwestern University, Evanston, Illinois 60208, USA}

\date{Received \today}

\begin{abstract}
A chimera state is a spatiotemporal pattern in which a network of identical coupled oscillators exhibits coexisting regions of asynchronous and synchronous oscillation. Two distinct classes of chimera states have been shown to exist: ``spots'' and ``spirals.''  Here we study coupled oscillators on the surface of a sphere, a single system in which both spot and spiral chimera states appear. We present an analysis of the birth and death of spiral chimera states and show that although they coexist with spot chimeras, they are stable in disjoint regions of parameter space.
\end{abstract}
\maketitle

% ********** Section **********
\section{Introduction}
Over the last half century, significant advances have been made in understanding the dynamics of coupled oscillators. Since the pioneering work of Winfree \cite{Winfree1967} and Kuramoto \cite{Kuramoto1975}, the nonlinear dynamics community has been able to use both analytical and numerical techniques to study the onset of synchronization and to explore other types of dynamics, including varying degrees of coherence and incoherence, in a broad class of oscillator networks. 

Before 2002, divergent behaviors such as incoherence and coherence were thought to result from heterogeneities.  However, Kuramoto and Battogtokh showed that even networks of identical oscillators could split into regions of coherence and incoherence \cite{Kuramoto2002}. Since this surprising discovery,  these ``chimera states'' have been reported in a vast array of network topologies \cite{Abrams2008,Laing2012_2,Martens2010,Shanahan2010,Yao2013,Zhu2014} including spatially embedded networks like a ring of oscillators \cite{Kuramoto2002,Abrams2004,Omelchenko2013}, a torus \cite{Panaggio2013,Omelchenko2012}, and a plane \cite{Shima2004,Martens2009,Laing2009_2,Gu2013}.   

Here we study the dynamics of Kuramoto oscillators on the surface a unit sphere $\mathbb{S}^2$. In this system, the phases $\psi(\rvec)$ are governed by
\begin{equation}
  \label{kuramoto}
  \frac{\partial\psi(\rvec)}{\partial t} = \omega-\int_{\mathbb{S}^2}G(\rvec,\rvecp )\sin (\psi(\rvec)-\psi(\rvecp)+\alpha)d\rvecp,
\end{equation}
where $G(\rvec,\rvecp)$ is a continuous coupling kernel.

We choose to study this system for several reasons. There are homeomorphisms (continuous deformations) from the sphere to many common closed two-dimensional surfaces embedded in three dimensions; spheres are topologically equivalent to all kinds of different surfaces with physical and biological relevance. Our results suggest that chimera states are likely to occur for oscillators on any orientable closed surface \cite{footnote}.

Furthermore, the sphere is a geometry in which both spot and spiral chimera states appear in very simple forms. These two unique dynamical patterns have yet to be connected from an analytical perspective. Spiral chimeras on the sphere show an intriguing similarity to patterns of activity displayed by the human heart during ventricular fibrillation \cite{Panfilov1998, Davidenko1992}.

\begin{figure*}
\center
\includegraphics[width=0.75\textwidth, trim = 0cm 0cm 0cm 0cm, clip=true]{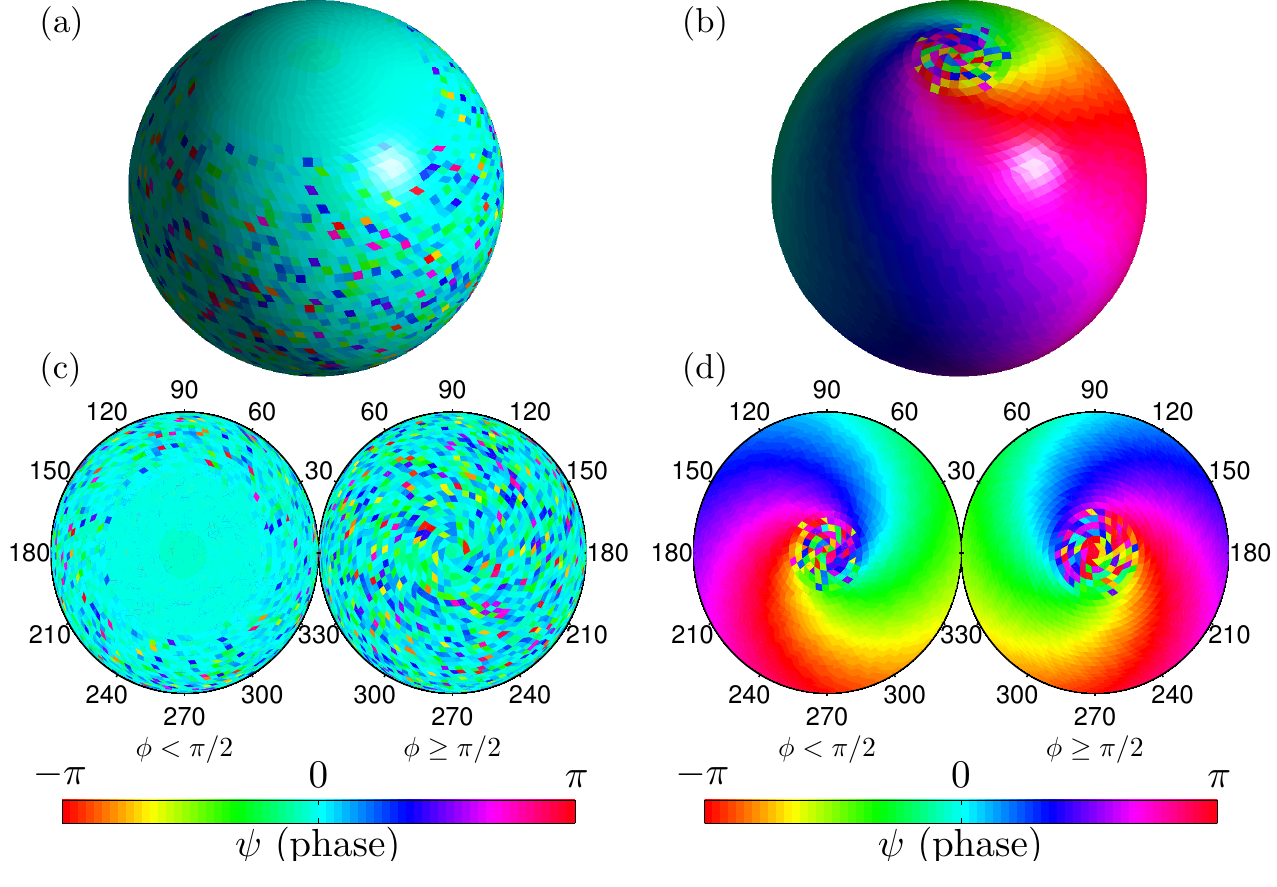}
\caption{(Color online) Examples of spot and spiral chimera states. Each panel displays the phases of 5000 oscillators corresponding to stable spots with $\Delta=0.947$, $\alpha=1.568$, and $\kappa=0.25$ [panels (a) and (c)] and spirals with $\Delta=0.478$, $\alpha=0.589$, and $\kappa=15$ [panels (b) and (d)].  Panels (a) and (b) display the sphere in three dimensions, while panels (c) and (d) display two-dimensional projections of a sphere from above and below. The phase of each oscillator is indicated by the color.}
\label{fig:spot_and_spiral_plot}
\end{figure*}

We analyze the dynamics of this system with near-global coupling and demonstrate the existence of spot and spiral chimera states in the perturbative limit. Then, using a variety of numerical and analytical techniques, we explore the role of the coupling {length} and phase lag $\alpha$ in determining the existence and stability of these unusual patterns (see Fig.~\ref{fig:spot_and_spiral_plot}).

% ********** Section **********
\section{Background} 
In two-dimensional spatially embedded networks of identical oscillators, two different classes of chimera states have been reported: spots and spirals. For spot chimeras, oscillators form spots of incoherence and coherence. In the coherent region, all oscillators share the same phase. For spiral chimeras, a region of incoherence is surrounded by a coherent region.  In the coherent region, the phases of the oscillators make a full cycle along any path around the incoherent spot. Thus the lines of constant phase resemble spiral arms around an implied phase singularity at the center of the incoherent region.  

Spot chimeras were discovered first. Their bifurcations \cite{Abrams2006,Panaggio2013} and stability \cite{Omelchenko2010,Omelchenko2013} have been studied extensively with near-global coupling. When $\alpha$ is near $\pi/2$, unstable and stable spot chimeras bifurcate off of the fully synchronized and drifting states respectively and then disappear due to a saddle node bifurcation. Thus they only exist near the Hamiltonian limit $\alpha=\pi/2$ \cite{Watanabe1993,Watanabe1994,Panaggio2014_2}. 

Spiral chimeras are not as well understood. They were reported by Shima and Kuramoto on an infinite plane \cite{Shima2004}. Their existence was confirmed analytically by Martens \etal \cite{Martens2009}, but their bifurcations and stability have not yet been studied from an analytical perspective. Numerical experiments have suggested that they are only stable when $\alpha$ is near 0 (a dissipative limit) and when the coupling is more localized \cite{Martens2009,Omelchenko2012}.

% ********** Section **********
\section{Analysis} 
We consider the special case where the coupling kernel is defined as
\begin{equation}
  \label{vm_kernel}
  G(\rvec,\rvecp) = \frac{\kappa}{4\pi\sinh {\kappa}}e^{\kappa \left( \rvec\cdot\rvecp \right)}.
\end{equation}
This is known as the Von-Mises-Fisher distribution \cite{Fisher1953} and represents the analog of a normal distribution on a sphere. The variance {(coupling length)} of this distribution is inversely related to the concentration parameter $\kappa$. As $\kappa\rightarrow\infty$,   $G(\rvec,\rvecp)\rightarrow \delta(\rvec-\rvecp)$ representing purely local coupling. When $\kappa\rightarrow 0$, $G(\rvec,\rvecp)\rightarrow 1/4\pi$ representing global coupling. We are interested in the role this concentration plays in the dynamics.

Following the approach of Kuramoto and Battogtokh \cite{Kuramoto2002}, we shift into a rotating frame with angular frequency $\Omega$ and define a complex order parameter, 
\begin{equation}
\label{orderparameter}
R(\rvec,t)e^{i\Psi(\rvec,t)}=\int_{\mathbb{S}^2}G(\rvec,\rvecp)e^{i\psi(\rvecp,t)}d\rvecp.
\end{equation}
Equation \eqref{kuramoto} can be rewritten in terms of the order parameter and the frequency difference $\Delta=\omega-\Omega$ revealing two types of stationary solutions: where $R(\rvec)\geq|\Delta|$, oscillators become phase-locked with a stationary phase, and where $R(\rvec)<|\Delta|$, they cannot become phase-locked but instead drift with a stationary phase distribution. In both the locked and drifting regions, these stationary solutions must satisfy a self-consistency equation 
\begin{equation}
  \label{self_c}
  R(\rvec)e^{i\Psi(\rvec)}=e^{i\beta}\int_{\mathbb{S}^2}G(\rvec,\rvecp)h(\rvecp)e^{i\Psi(\rvec',t)}d\rvecp
\end{equation}
where $h(\rvec)=\frac{\Delta -\sqrt{\Delta^2-R^2(\rvec)}}{R(\rvec)}$ and $\beta=\pi/2-\alpha$. 

To reduce the dimensionality of this system, we parametrize the surface of the sphere using the mathematical convention for spherical coordinates
\begin{equation*}
  \rvec=\cos{\theta}\sin{\phi}\hat{\textbf{i}}+\sin{\theta}\sin{\phi}\hat{\textbf{j}}+\cos{\phi}\hat{\textbf{k}},
\end{equation*} 
where $\theta$ represents the azimuthal angle and $\phi$ represents the polar angle and restrict our search to solutions of the form
\begin{subequations}\label{self_c_ans}
\begin{align}
  R(\rvec)   &= A(\phi) \\
  \Psi(\rvec)&= B(\phi)+N\theta
\end{align}
\end{subequations} 
where $N$ is an integer. These solutions correspond to rotationally symmetric spots ($N=0$), simple spirals ($N=1$), and higher order spirals ($N>1$). Under this restriction, Eq.~\eqref{self_c} becomes
\begin{align}
  \label{new_op}
  A(\phi)e^{iB(\phi)} &= e^{i\beta}\int_{0}^{\pi}K_N(\phi,\phip)h(\phip)e^{iB(\phip)}\sin{\phip} d\phip
\end{align}
where 
\begin{align*}
  K_N(\phi,\phip) &= \int_{0}^{2\pi}G(\theta,\phi,\thetap,\phip)e^{iN(\thetap-\theta)} d\thetap\\
                  &= \frac{\kappa}{2\sinh{\kappa}}e^{\kappa\cos{\phi}\cos{\phip}}I_N(\kappa\sin{\phi}\sin{\phip})~, 
\end{align*}
where $I_N$ is the $N^\text{th}$ order modified Bessel function of the first kind. Note that $K_N$ is independent of $\theta$. Equation \eqref{new_op} is a complex nonlinear integral eigenvalue problem. Solving explicitly for $A(\phi)$, $B(\phi)$, and $\Delta$ is not possible in general, and solutions may not exist for all $\beta,\kappa$.

% ********** Subsection **********
\subsection{Near-global coupling}\label{sec:global} 
With near-global coupling ($\kappa\ll 1$), the coupling kernel can be approximated to leading order in $\kappa$ by 
\begin{multline*}
  G(\theta,\phi,\thetap,\phip)= \frac{1}{4\pi} \{1+\kappa[\cos{\phi}\cos{\phip}\\+\cos{\left(\theta-\thetap\right)}\sin{\phi}\sin{\phip}]\}.
\end{multline*}
Substitution of this coupling kernel into Eq.~\eqref{self_c} reveals that the order parameter must have the following form
\begin{multline}
  \label{self_c_2}
  R(\rvec)e^{i\Psi(\rvec)}= c+\kappa d_1\cos{\theta}\sin{\phi}\\+\kappa d_2\sin{\theta}\sin{\phi}+\kappa d_3\cos{\phi},
\end{multline}
where $\langle f(\rvecp) \rangle = \frac{1}{4\pi}\int_{\mathbb{S}^2} f(\rvecp)d\rvecp$ and
\begin{subequations}\label{sc_alg}
\begin{align}
  c   &= e^{i\beta}\langle h(\rvecp)e^{i\Psi(\rvecp)}\rangle \label{sc_alg_c}\\
  d_1 &= e^{i\beta}\langle h(\rvecp)e^{i\Psi(\rvecp)}\cos{\thetap}\sin{\phip}\rangle \label{sc_alg_d1}\\
  d_2 &= e^{i\beta}\langle h(\rvecp)e^{i\Psi(\rvecp)}\sin{\thetap}\sin{\phip}\rangle \label{sc_alg_d2}\\
  d_3 &= e^{i\beta}\langle h(\rvecp)e^{i\Psi(\rvecp)}\cos{\phip}\rangle. \label{sc_alg_d3}
\end{align}
\end{subequations}
Note that the derivation of Eq.~\eqref{sc_alg} does not rely on Eq.~\eqref{self_c_ans}. We now focus on the case where $N=0$. 

% ********** Subsubsection **********
\subsubsection{Spot chimeras}
\begin{figure}[t]
  \center
  \includegraphics[width=0.9\columnwidth]{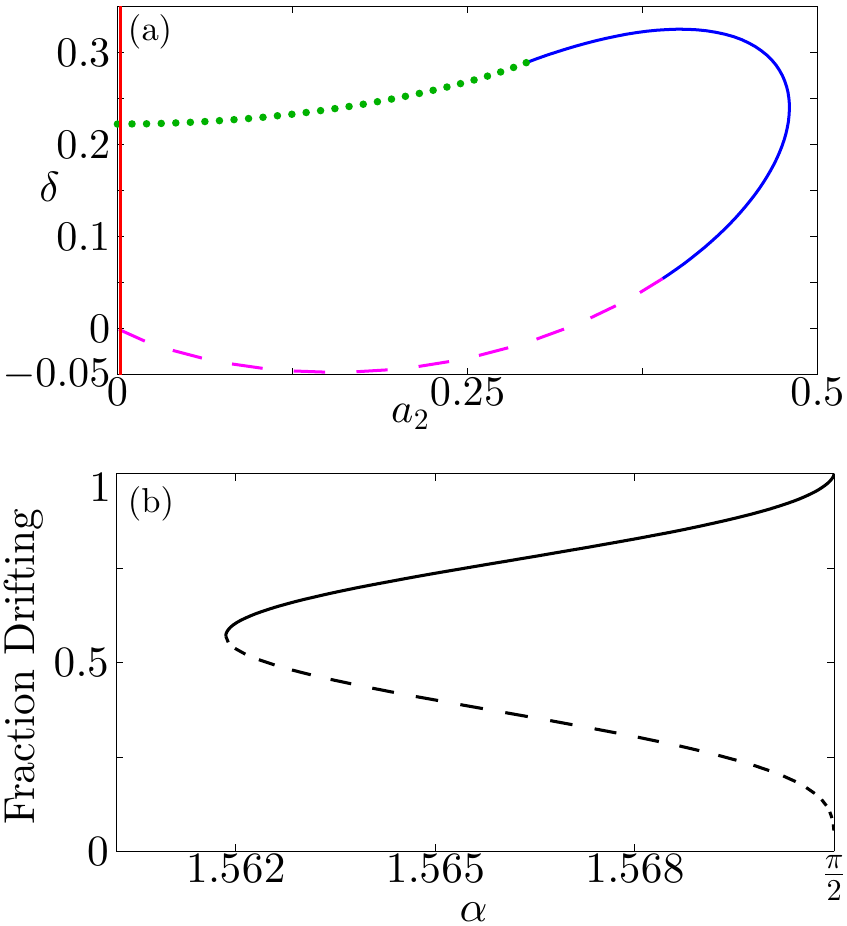}
  \caption{(Color online) Solutions to Eq.~\eqref{self_c_10}.  Panel (a) displays parameters $a_2$ and $\delta=\Delta_2-c_2$ [see Eq.~\eqref{scaling} for the definitions]. The green (dotted) curve corresponds to solutions with spatially modulated drift. The blue (solid) curve corresponds to stable spot chimeras.  The magenta (dashed) curve corresponds to unstable spot chimeras. The red (solid) curve along the vertical axis represents uniform solutions.  Panel (b) displays the fraction of oscillators in the drifting region as a function of $\alpha$ (where $\kappa=0.1$). The solid curve represents stable chimera states and the dashed curve represents unstable chimera states. }
  \label{fig:spot_res}
\end{figure}

When $N=0$, $R$ and $\Psi$ (and therefore $h$) depend only on $\phi$, so integration with respect to $\thetap$ is possible giving $d_1=d_2=0$. Thus the order parameter takes the form  
\begin{equation}  \label{self_c_8}
  R(\rvec)e^{i\Psi(\rvec)}=A(\phi)e^{iB(\phi)}=c+d\cos{\phi}
\end{equation}
where $d=\kappa d_3$. This yields two equations:
%\begin{subequations}\label{self_c_9}
\begin{align*}
  c &= \frac{1}{2}e^{i\beta}\int_0^{\pi}\frac{\Delta-\sqrt{\Delta^2-\left|c+d\cos{\phi}\right|^2}}{c^*+d^*\cos{\phi}}\sin{\phip}  d\phip \\% \label{self_c_9a}~, \\
  d &= \frac{1}{2}\kappa e^{i\beta}\int_0^{\pi}\frac{\Delta-\sqrt{\Delta^2-\left|c+d\cos{\phi}\right|^2}}{c^*+d^*\cos{\phi}}\cos{\phip}\sin{\phip} d\phip %\label{self_c_9b}.
\end{align*}
where $^*$ denotes complex conjugation.
%\end{subequations}

Motivated by the results from \cite{Abrams2006,Panaggio2013} we look for solutions that scale like
\begin{subequations}\label{scaling}
\begin{align}
  \beta  &= \beta_1\kappa\\
  c      &\sim 1+c_1\kappa+c_2\kappa^2 \\
  d      &\sim (a_2+ib_2)\kappa^2  \\
  \Delta &\sim 1+\Delta_1\kappa+\Delta_2\kappa^2.
\end{align}
\end{subequations}
Expanding in $\kappa$ and defining $\delta=\Delta_2-c_2$ we obtain the following conditions at leading order:
\begin{subequations}\label{self_c_10}
\begin{align}
  c_1        &= \frac{\sqrt{2}}{3}\left[\left(\delta-a_2\right)^{3/2}-\left(\delta+a_2\right)^{3/2}\right]+i\beta_1~, \label{self_c_10a}\\
  a_2 + ib_2 &= \frac{\sqrt{2}}{a_2^2}\bigg[\frac{2\delta}{3}\left(\left(\delta-a_2\right)^{3/2}-\left(\delta+a_2\right)^{3/2}\right)\nonumber\\&-\frac{2}{5}\left(\left(\delta-a_2\right)^{5/2}-\left(\delta+a_2\right)^{5/2}\right)\bigg]~.    \label{self_c_10b}
\end{align}
\end{subequations}

Note that $\Delta_1=c_1$ was required to satisfy the equations at $\mathcal{O}(\sqrt{\kappa})$. The real part of Eq.~\eqref{self_c_10a} depends only on $\delta$ and $a_2$.  Thus we can fix $\delta$, solve for $a_2$, and then compute the other unknowns directly.  The parameters $a_2$ and $\delta$ determine the variation in the order parameter and the size of the drifting region. Using MatCont \cite{Dhooge2003}, a numerical continuation software package for MATLAB, we find the solutions to Eq.~\eqref{self_c_10} and display them in Fig.~\ref{fig:spot_res}(a). Figure \ref{fig:spot_res}(b) shows the fraction drifting as a function of $\alpha$. These solutions resemble the spot solutions observed in Refs.~\onlinecite{Abrams2006,Panaggio2013} in that unstable spot chimeras bifurcate off of a phase-locked state and stable spot chimeras bifurcate off of a modulated drift state. At a critical value of $\beta = \frac{4}{1594323}(188\sqrt{10}+505)^{3/2} \kappa \approx 0.0915\kappa$, the chimera states disappear due to a saddle node bifurcation (see Appendix \ref{app_snb}).  This explains the change in stability observed in panel (b). The stability of these solutions was confirmed via numerical integration of Eq.~\eqref{kuramoto}. 

% ********** Subsubsection **********
\subsubsection{Spiral chimeras} 
\begin{figure}[b]
\center
\includegraphics[width=0.9\columnwidth, trim = 0cm 0cm 0cm 0cm,clip=true]{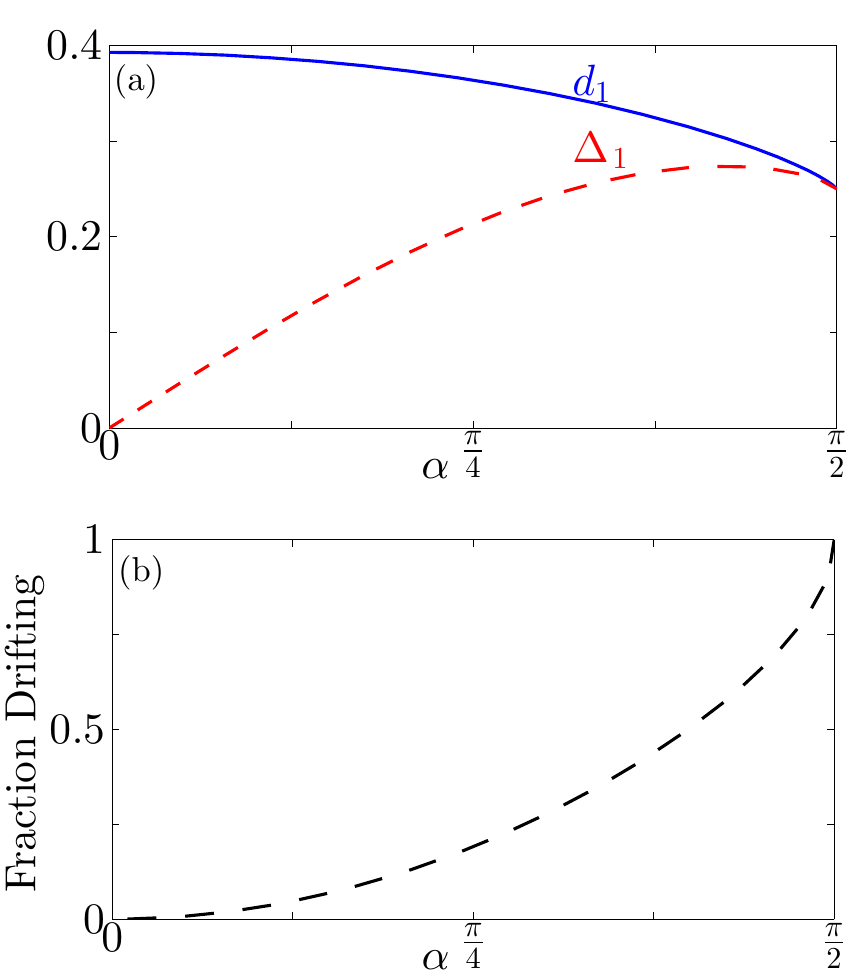}
\caption{(Color online) Solutions to Eq.~\eqref{self_c_7}.  Panel (a) displays the values of $d_1$ (blue solid) and $\Delta_1$ (red dashed) as a function of $\alpha$. Panel (b) displays the fraction of oscillators in the drifting region as a function of $\alpha$. }
\label{fig:smallk_spiral_sc}
\end{figure}

No higher order spirals ($N>1$) occur to lowest order in $\kappa$ because all terms in Eq.~\eqref{sc_alg} integrate to 0. For $N=1$, on the other hand, $c=d_3=0$ and $d_2=id_1$, so Eq.~\eqref{self_c_2} yields an order parameter of the form
\begin{equation}
  \label{self_c_3}
  R(\rvec)e^{i\Psi(\rvec)}=A(\phi)e^{iB(\phi)}e^{i\theta}=\kappa d_1 \sin{\phi} e^{i\theta}.
\end{equation}
Without loss of generality, we can define the argument $\Psi(\rvec)$ to be 0 along the half plane $\theta=0$ and making $d_1$ real. This implies that $A(\phi)=\kappa d_1\sin{\phi}$ and $B(\phi)=0$.  Defining $\Delta_1=\Delta/\kappa$, substituting this result into Eq.~\eqref{self_c}, and integrating with respect to $\theta$ yields   
\begin{equation}
  \label{self_c_6}
  d_1^2=\frac{e^{i\beta}}{4} \int_0^{\pi} \left(\Delta_1 -\sqrt{\Delta_1^2-d_1^2 \sin^2\phip}\right)\sin{\phip}d\phip.
\end{equation}
Note that chimera states only appear if $d_1>\left|\Delta_1\right|$. [For $d_1\leq\left|\Delta_1\right|$, Eq.~\eqref{self_c_6} can only be satisfied for $\beta=0$.]  Integrating with respect to $\phi$, this simplifies to
\begin{multline}
  \label{self_c_7}
  d_1^2=\frac{ e^{i\beta}}{4} \bigg[\frac{1}{2d_1}\left(\Delta^2-d_1^2\right)\left(\ln{\left(\frac{d_1-\left|\Delta_1\right|}{d_1+\left|\Delta_1\right|}\right)}+i\pi\right)\\+\left(2\Delta_1-\left|\Delta_1\right|\right)\bigg].
\end{multline}
where $\ln$ refers to the principal branch of the natural logarithm. 

Solving Eq.~\eqref{self_c_7} reveals that like spot chimeras, spiral chimeras represent a link between coherence and incoherence (see Fig.~\ref{fig:smallk_spiral_sc}).  We find that when $\alpha=0$, $d_1=\pi/8$ and $\Delta_1=0$.  This leads to a fully locked solution in which the phases depend only on the azimuthal variable $\theta$ (a ``beachball'' pattern). As $\alpha$ increases from 0, incoherent spiral cores are born at the poles of the sphere and grow until $\alpha=\pi/2$. When $\alpha=\pi/2$, $d_1=\Delta_1=1/4$, and the sphere is fully incoherent. Although these spirals resemble the spirals reported in Refs.~\onlinecite{Martens2009,Gu2013,Omelchenko2012}, numerical experiments suggest that these states are unstable.  They only seem to gain stability when coupling is more localized.

% ********** Subsection **********
\subsection{Localized coupling}\label{sec:local}
To search for stable spiral chimera states, we now explore the dynamics when $\kappa$ is not small. With highly localized coupling, the effects of curvature are negligible, and the sphere can be approximated locally as a plane.  Martens \etal showed that, on an infinite plane, spiral chimera states appear when $\alpha$ is small.  Motivated by these findings, we consider the limit where $\alpha \ll 1$ and assume the following scalings
\begin{subequations}\label{scalings}
\begin{align*}
  \Delta  &= \Delta_1 \alpha+\mathcal{O}(\alpha^2)\\%\label{scalings_Delta} \\
  A(\phi) &= A_0(\phi)+A_1(\phi) \alpha+\mathcal{O}(\alpha^2)\\%\label{scalings_A} \\
  B(\phi) &= B_1(\phi) \alpha+\mathcal{O}(\alpha^2)~. %\label{scalings_Psi}. 
\end{align*}
\end{subequations}
Expanding Eq.~\eqref{new_op} in $\alpha$ to leading order yields
\begin{align*}
  A_0(\phi)=\int_{0}^{\pi}K_1(\phi,\phip)\sin{\phip} d\phip~,
\end{align*}
which can be integrated numerically to find $A_0(\phi)$. To $\mathcal{O}(\alpha)$ we find that 
\begin{align*}
  A_1(\phi) &= 0 \\
  B_1(\phi) &= \int_{0}^{\pi}\frac{K_1(\phi,\phip)}{A_0(\phi)}\left(B_1(\phip) + \frac{\Delta_1}{A_0(\phip)}\right)\sin{\phip} d\phip-1.
\end{align*}
Thus $B_1(\phi)$ and $\Delta_1$ satisfy an inhomogeneous Fredholm equation of the second kind which can be solved numerically.  This asymptotic approach decouples the magnitude of the order parameter from its argument making it possible to solve for each separately.  It also allows a nonlinear equation to be approximated by a series of linear equations and can be used to find higher order approximation to $A$ and $B$ as well.
  
These results can be used to estimate the size of the incoherent region at the center of the spiral. To see this, let $\phi=\phi_B$ represent the boundary between the locked and drifting regions. To order $\alpha$, the boundary satisfies
\begin{equation}
\label{boundary}
  A_0(\phi_B)=\alpha\left|\Delta_1\right|.
\end{equation}
There are two solutions to this equation that are symmetric about the equator, one with $\phi_B\approx 0$ and one with $\phi_B\approx \pi$. To find the size of the incoherent region with $\phi_B\approx 0$, we expand $A_0$ about $\phi=0$,     
\begin{align}
\label{boundary_series}
\begin{split}
  A_0(\phi_B) &\sim A_0(0)+\phi_B A_0^{\prime}(0) \\
              &\sim 0 + \phi_B \int_{0}^{\pi}\frac{\partial}{\partial \phi}K_1(0,\phip)\sin{\phip} d\phip.
\end{split}
\end{align} 
Substituting Eq.~\eqref{boundary} into Eq.~\eqref{boundary_series}, integrating with respect to $\phip$, and solving for $\phi_B$ yields 
%&=\frac{\alpha\left|\Delta_1\right|}{\int_{0}^{\pi}\frac{\partial}{\partial \phi}K(0,\phip|1)\sin{\phip} d\phip)}\\
\begin{align}
\label{phib}
  \phi_B &=\frac{4\sinh{\kappa}}{\pi \kappa I_1(\kappa)}\alpha\left|\Delta_1\right|.
\end{align} 
Thus, near the birth of the chimera state the size of the incoherent region grows with $\alpha$.   {Since $\Delta_1$ is $\kappa$ dependent but its scaling with $\kappa$ is unknown, the dependence of the size of the incoherent region on $\kappa$ cannot be determined using Eq.~\eqref{phib}. However, given a numerical solution for $A(\phi)$ and $\Delta$, the value of $\phi_B$ is readily apparent and the size of the drifting region can be easily calculated (see Fig.~\ref{fig:fracs} in Appendix \ref{app_ns}).}

% ********** Section **********
\section{Numerical continuation}
The asymptotic approximations discussed in Secs.~\ref{sec:global} and \ref{sec:local} are only valid when $\alpha \ll 1$ or $\kappa \ll 1$.  To explore other regions of parameters space, we used MatCont \cite{Dhooge2003} for numerical continuation.  This software package uses Newton's method to allow the user to continue equilibria of systems of ordinary differential equations (ODEs) and to detect bifurcations. 

For numerical continuation, we defined $w(\phi)=(A(\phi)/\Delta)e^{iB(\phi)}$ and rewrote Eq.~\eqref{new_op} as 
\begin{equation}
  w(\phi)\Delta=e^{i\beta}\int_{0}^{\pi}\frac{K_N(\phi ,\phip)w(\phip)\sin{\phip}}{1 +\sqrt{1-\left|w(\phip)\right|^2}} d\phip.
  \label{op_w} 
\end{equation} 
The known spiral solutions for $w(\phi)$ are nearly sinusoidal, therefore it is natural to represent them by a Fourier sine series
\begin{equation*}
  w(\phi)=\sum_{n=1}^{\infty} (x_n+iy_n)\sin(n\phi)
\end{equation*}  
where 
\begin{align*}
  x_n &= \frac{2}{\pi}\int_0^{\pi}\textnormal{Re}(w(\phi))\sin(n\phi) d\phi,\\
  y_n &= \frac{2}{\pi}\int_0^{\pi}\textnormal{Im}(w(\phi))\sin(n\phi) d\phi.
\end{align*}
In Fourier space, it is straightforward to show that the fixed points of 
\begin{subequations}
\begin{align}
\label{eq21a}
  x_n^{\prime}    &= \textnormal{Re} \left(\frac{e^{i\beta}}{\Delta} \int_0^{\pi}\frac{K_1^n(\phip)w(\phip)\sin{\phip}}{1 +\sqrt{1-\left|w(\phip)\right|^2}}d\phip \right)-x_n,\\  
 \label{eq21b}  
  y_n^{\prime}    &= \textnormal{Im} \left(\frac{e^{i\beta}}{\Delta} \int_0^{\pi}\frac{K_1^n(\phip)w(\phip)\sin{\phip}}{1 +\sqrt{1-\left|w(\phip)\right|^2}} d\phip \right)-y_n,\\ 
\label{tmp}  
  \Delta^{\prime} &= \max\left(\textnormal{arg }w(\phi)\right),
\end{align}
\label{matcont_f2}
\end{subequations}
where $\textnormal{Re} $ and $\textnormal{Im}$ denote real and imaginary parts and 
\begin{equation*}
  K_1^n(\phip)=\frac{2}{\pi}\int_0^{\pi}K_1(\phi,\phip)\sin(n\phi) d\phi
%K(\phi,\phip|N)=\sum_{n=1}^{\infty} K_1^n(\phip)\sin(n\phi) \textnormal{ and } 
\end{equation*}
represents solutions to Eq.~\eqref{op_w}. Equation \eqref{tmp} was imposed to eliminate the extra degree of freedom in Eq.~\eqref{op_w} (invariance under rotations $w\rightarrow w e^{ir}$).

% ********** Section **********
\section{Results}
% ********** Subsection **********
\subsection{Existence}
\begin{figure*}
  \center
  \includegraphics[width=0.75\textwidth, trim = 0cm 0cm 0cm 0cm,clip=true]{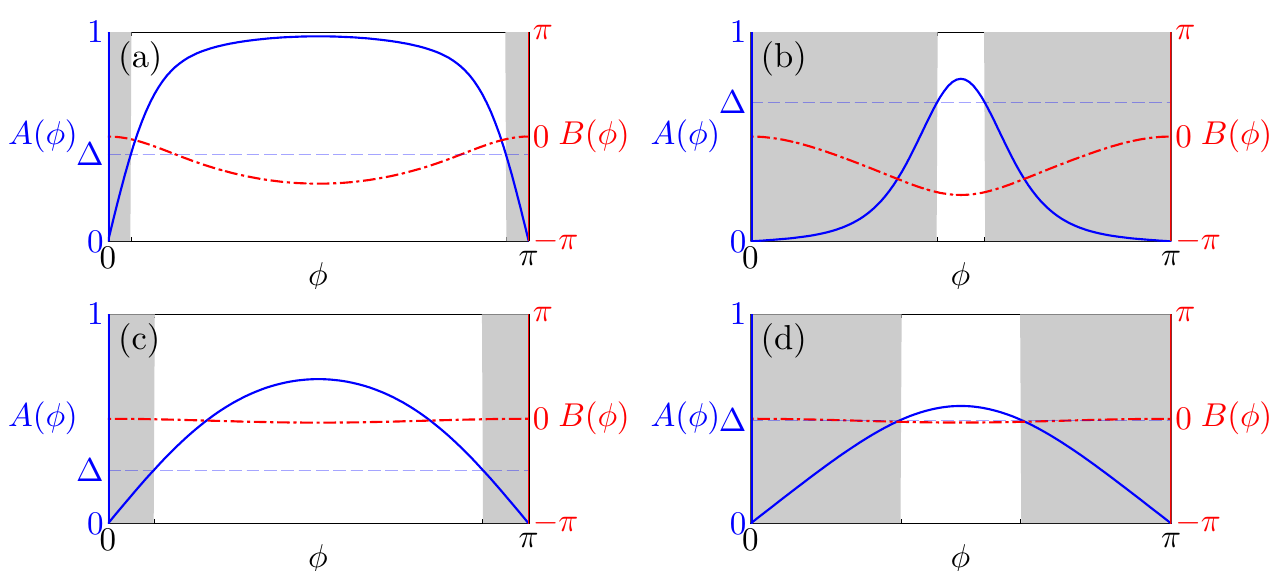}
  \caption{(Color online) Typical solutions to Eq.~\eqref{new_op}. Panels (a)-(d) display $A(\phi)$ and $B(\phi)$ for $\alpha=0.47$ and $\kappa=25$, $\alpha=1.30$ and $\kappa=25$, $\alpha=0.47$ and $\kappa=2.5$, and $\alpha=1.30$ and $\kappa=2.5$ respectively. The blue (solid) curve represents $A(\phi)$ and the red (dash-dotted) represents $B(\phi)$. Gray boxes denote the drifting regions $[0,\phi_B)$ and $(\pi-\phi_B,\pi]$.} 
  \label{fig:op_sols}
\end{figure*}

\begin{figure}[b]
\includegraphics[width=0.45\textwidth, trim = 0cm 0cm 0cm 0cm, clip=true]{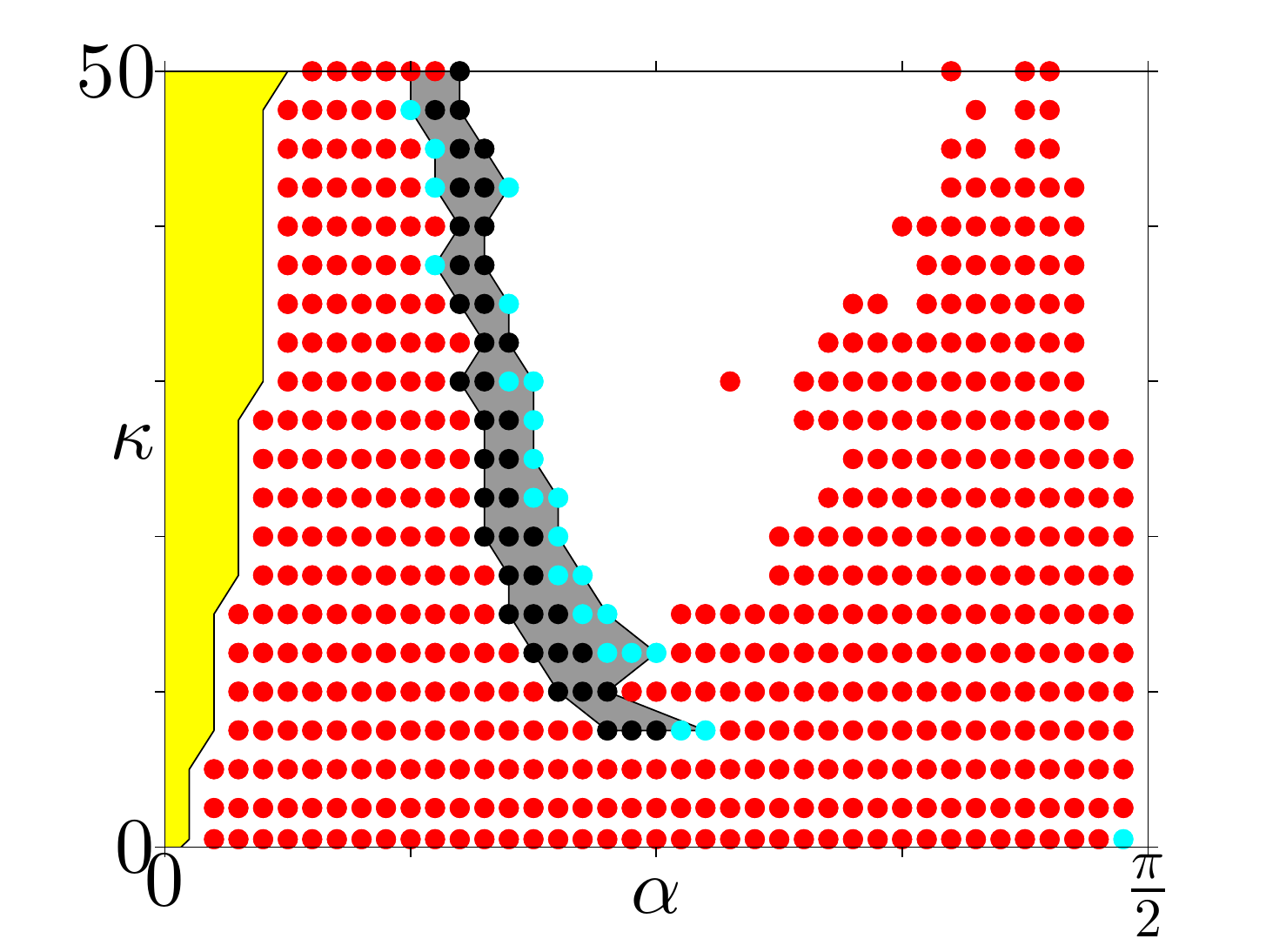}
\caption{(Color online) Existence and stability of spiral chimera states. Dots represent points where spiral chimera states satisfying  Eq.~\eqref{new_op} exist.  Red (dark) points correspond to chimeras that are unstable in numerical simulations. Black points correspond to chimeras that appear to be stable after integrating Eq.~\eqref{kuramoto} for 5000 units of time. Cyan (light) points correspond to chimeras where the final state is a spiral but shows noticeable deviation from the initial condition.  In the yellow shaded region (with $\alpha$ near 0) chimera states exist but have too few incoherent oscillators to reliably assess stability (see Appendix \ref{app_ns}).  All of the spiral chimeras that appear to be stable are contained in the narrow gray strip (with intermediate values of $\alpha$).}
\label{fig:spiral_stability}
\end{figure}

The results from numerical continuation of spiral chimeras are displayed in Figs.~\ref{fig:op_sols} and \ref{fig:spiral_stability}. We find that spiral chimera states satisfying Eqs.~\eqref{op_w} and \eqref{new_op} continue to exist for $\kappa \gtrsim \mathcal{O}(1)$. Near $\alpha=0$ and $\alpha=\pi/2$, we were able to continue these solutions indefinitely in $\kappa$.  For intermediate values of $\alpha$ the numerical continuation fails prematurely at $\kappa_{\textnormal{max}}<50$. Attempts at continuing beyond this point by increasing the number of Fourier coefficients retained, refining the grid for numerical integration, and continuing using alternative methods (see Appendix \ref{app_nc}) yielded incremental increases in $\kappa_{\textnormal{max}}$. This suggests that the failure of convergence is due to narrowing of the basin of attraction for the numerical solution and increasingly sharp transitions in the shape of the solutions, however we cannot rule out a failure of existence due to a bifurcation.

Qualitatively these spirals resemble the ones observed for $\kappa \ll 1$. They are symmetric with respect to reflections about the equator. However, instead of straight spiral arms (lines of constant phase) where $B(\phi)\approx 0$ (see Fig.~\ref{fig:op_sols}, bottom panels), these solutions have curved spiral arms with $B(\phi) \neq 0$ (see Fig.~\ref{fig:op_sols}, top panels). The fraction of oscillators drifting is zero for $\alpha=0$ and increases with $\alpha$ until the entire sphere is incoherent when $\alpha=\pi/2$. See Fig.~\ref{fig:spot_and_spiral_plot} for an example.  

Spot chimeras appear to exist for arbitrary $\kappa$ and seem qualitatively similar to solutions for $\kappa \ll 1$ (see Appendix \ref{app_scwlc}).

% ********** Subsection **********
\subsection{Stability}
In order to test the stability of these solutions, we approximated Eq.~\eqref{kuramoto} by selecting 5000 points uniformly distributed on the surface of a sphere \cite{Thomsen2007,Rakhmanov1994}, generating initial conditions consistent with the order parameters obtained through numerical continuation (see Appendix \ref{app_gic}), and integrating for 5000 units of time (10-1000 cycles of the locked oscillators, depending on the values of $\alpha$ and $\kappa$).  After this interval, if the final state possessed a phase distribution that was nearly identical to the initial state (except for possible drifting of the incoherent region)  we classified the chimera state as stable.  

We observed a narrow strip with stable chimeras extending down to $\alpha\approx0.85$ and $\kappa\approx 7.5$. We believe that, to conform with the planar case explored in Ref.~\onlinecite{Martens2010}, this strip is likely to originate from $\alpha\approx0$ and $\kappa \gg 1$ (in this limit the coupling is so localized that the curvature of the sphere becomes irrelevant). Near the boundaries of this strip, solutions remained close to the initial condition for most of the integration time before evolving toward a fully coherent state or spiral pattern without an incoherent region, suggesting that spiral chimera states outside of the strip were unstable.  At the moment we have no analytical explanation for the observed changes in stability. One possibility is that the states we refer to as stable are actually just very long-lived transients. However, that raises the question of why this particular strip would have dramatically longer transient times than neighboring regions of parameter space.  Another possibility is that stability changes due to some as of yet unidentified bifurcation.  This bifurcation cannot be due to the presence of a spot chimera because of the topological differences and the fact that spot chimeras do not exist near the boundaries of this region, but it could be due to other equilibrium spiral patterns that only satisfy ansatz \eqref{self_c_ans} at the bifurcation point.

% ********** Section **********
\section{Discussion and conclusions}
This work demonstrates the existence of both spot and spiral chimera states on the surface of a sphere. We find that both spirals and spots represent links between coherence and incoherence. In agreement with previous results, when coupling is nearly global, spot chimeras only exist near the Hamiltonian limit ($\alpha=\pi/2$) whereas spiral chimeras exist for all values of $0 \leq \alpha \leq \pi/2$. For more localized coupling, numerical results suggest that both types of chimera states continue to exist, but that they have disjoint regions of stability.
{A puzzling apparent failure of existence of chimera states for localized coupling and intermediate phase lags ($\alpha \approx \pi/4$) remains to be explained.}

More broadly, we have demonstrated that the surface of a sphere provides an interesting testbed for assessing the properties of chimera states---one in which analogs of many previously reported chimera states exist. Although the underlying cause is the same, this topology leads to visually distinct patterns from other two dimensional systems---on a plane, single spirals appear, whereas, on a torus, spirals only appear in multiples of 4 and on a sphere spirals appear in pairs.  The result that both spiral and spot chimera states occur over a wide range of parameter values in these systems suggests that chimera states may be possible in any network of non-locally coupled oscillators on a closed, orientable surface.  In particular, the topological resemblance of a sphere to real-world systems makes this geometry potentially valuable for applications to naturally occurring biological oscillatory networks (e.g. the human heart and brain, where chimera states could be associated with dangerous ventricular fibrillation or epileptic seizure states).

\begin{acknowledgments}
The authors would like to thank C.~Laing for useful conversations.
\end{acknowledgments}

%% Start appendices
%\setcounter{section}{0}
%\setcounter{equation}{0}
%\setcounter{figure}{0}
%\renewcommand\thesection{APPENDIX \Alph{section}\setcounter{figure}{0}\setcounter{equation}{0}}
%\renewcommand\thesubsection{\arabic{subsection}}
%\renewcommand\theequation{\Alph{section}\arabic{equation}}
%\renewcommand\thefigure{\Alph{section}\arabic{figure}}

\appendix

% ********** Section **********
\section{Computing saddle-node bifurcation} \label{app_snb}

The saddle-node bifurcation with respect to $\beta$ (or, equivalently, $\alpha$) visible in Fig.~2(b) is straightforward to compute numerically from Eqs.~(11).  To compute an analytical form for the critical $\beta_1$, however, we proceeded as follows:
\begin{itemize}
\item Isolate $\delta$ in the imaginary part of (11a), then eliminate $\delta$ by plugging into the real part of (11b) to get a function $f(a_2, \beta_1) = 0$.
\item Find the maximum $\beta_1$ for which a solution exists by differentiating $f=0$ with respect to $a_2$ and imposing $d \beta_1/d a_2 = 0$, then solving for $\beta_1(a_2)$.
\item Plug in the result to get $f(a_2,\beta_1(a_2))=0$ and solve for $a_2$ to get $a_2^{(crit)} = \frac{86}{2187} \sqrt{10} + \frac{580}{2187}$.
\item Plug $a_2 = a_2^{(crit)}$ into $\beta_1(a_2)$ to get $\beta_1^{(crit)} = \frac{4}{1594323}\left(188 \sqrt{10} + 505 \right)^{3/2}$.
\end{itemize}  
Note that in each step itemized here significant simplification may be required to obtain a suitably concise result.

% ********** Section **********
\section{Numerical continuation} \label{app_nc}

The nonlinearity of Equation \eqref{new_op} made it unlikely that numerical methods would converge to the correct solution without an accurate initial guess.  So, we began with the solutions for $A(\phi)$, $B(\phi)$, and $\Delta$ derived for $\kappa \ll 1$ and $\alpha \ll 1$ and then implemented a variety of algorithms in order to numerically continue spiral chimera states over the parameter space $0 \leq \alpha \leq \pi/2$ and $0 \leq \kappa$.

% ********** Subsection **********
\subsection{Iterative method}
The simplest approach we implemented was a naive iterative method. Eq.~\eqref{op_w} has the form $\Delta w = f(w)$. Given an initial guess for the solution to Eq.~\eqref{op_w} $w_0(\phi)$ (in practice we used a discrete set of $\phi$ values), we updated our solution as follows:\\\\
\noindent \textit{Step 1:}\\Define $f_{n+1}(\phi)=e^{i\beta}\int_{0}^{\pi}K_N(\phi ,\phip)\frac{w_n(\phip)}{1 +\sqrt{1-\left|w_n(\phip)\right|^2}}\sin{\phip} d\phip$.

\noindent \textit{Step 2:}\\ Choose $\Delta_{n+1}$ to minimize $E=\left|\frac{1}{\Delta_{n+1}}f_{n+1}(\phi)-w_{n}(\phi)\right|$.

\noindent \textit{Step 3:}\\Update $w_{n+1}(\phi)=\frac{1}{\Delta_{n+1}}f_{n+1}(\phi)$.\\\\
To carry out numerical continuation using this heuristic scheme, we selected a known solution as a starting point made a small change to one of the parameters, and then iterated until the residual $r_{n+1}=w_{n+1}(\phi)-w_{n}(\phi)$ was small. This method is not guaranteed to converge, but for a starting point sufficiently close to the true solution and steps that were sufficiently small, it did allow us to identify solutions for new ranges of $\alpha$ and $\kappa$.

% ********** Subsection **********
\subsection{Optimization method} \label{app_om}
In order to improve upon the above, we implemented a second similar method.  We again started with a known solution and made a small change to either $\alpha$ or $\kappa$.  Then, we utilized a quasi-Newton method (as implemented in the MATLAB function fminunc) to find the values of $w(\phi)$ and $\Delta$ that minimized the error:
\begin{equation}
  E=\left| w(\phi)\Delta-e^{i\beta}\int_{0}^{\pi}\frac{K_N(\phi ,\phip)w(\phip)\sin{\phip}}{1 +\sqrt{1-\left|w(\phip)\right|^2}} d\phip\right|.\label{error}
\end{equation} 
This method seemed to be more stable than the iterative approach, but it was more computationally intensive.

% ********** Subsection **********
\subsection{Improvements and limitations}
Both of the above methods used zeroth order extrapolation to generate a starting point for the next set of parameter values. To improve upon this method we also used first order extrapolation to generate the next starting point.  For example, to continue in $\kappa$, given $\kappa_1<\kappa_2$ and their associated solutions for $\Delta$ and $w(\phi)$, we used a linear approximation to generate a guess at $\kappa_3>\kappa_2$. This guess was then used as a starting point for the above methods.  Although the above approaches did yield marginal gains in exploring $\alpha$ vs.~$\kappa$ space, ultimately the time and memory demands were far too large to adequately explore the domain of interest due to the small step sizes required for convergence. 

% ********** Subsection **********
\subsection{MatCont}
We found that MatCont was the most effective method for numerically continuing spiral chimera states. Our first attempt at writing Eq.~\eqref{new_op} as a system of ODEs, the input format required by MatCont, used a discretized version of Eq.~\eqref{op_w} on a uniform grid of 101 points between $0\leq \phi\leq \pi$. Unfortunately, the algorithm was unable to identify an appropriate search direction for continuation.

Instead we represented $w(\phi)$ as a Fourier sine series as described in the main text.  For $\kappa \ll 1$, $w(\phi)$ is sinusoidal, thus only the first Fourier coefficient is nonzero.  As $\kappa$ increases subsequent terms become more important.  For $\kappa < 50$, 
\begin{equation*}
\bigg|\bigg|\sum_{n=1}^{\infty} (x_n+iy_n)\sin(n\phi)-\sum_{n=1}^{16} (x_n+iy_n)\sin(n\phi)\bigg|\bigg| \lesssim \mathcal{O}(10^{-4})
\end{equation*} 
So, we truncated the series after 16 terms (inclusion of higher frequency modes does not significantly change the results). We then verified the accuracy of these solutions by substituting them directly into equation \eqref{op_w}. The integrals in Eqs.~\eqref{eq21a} and \eqref{eq21b} were evaluated using Simpson's rule with 101 grid points. We terminated continuation when the change in $\kappa$ over 10 steps was less than $10^{-5}$.    

% ********** Subsection **********
\subsection{Interpolation and refinement}
The endpoints ($\kappa_{\textnormal{max}}$) of the curves obtained from MatCont were somewhat irregular. To address this, we compiled all of the solutions from MatCont and used spline interpolation and extrapolation to generate guesses for missing solutions.  Then, using the optimization method described above we refined the guesses until the optimization scheme terminated and computed the error for these new points using Eq.~\eqref{error}. If the norm of the error was less than $10^{-4}$, we accepted the result as a solution to Eq.~\eqref{op_w}.  This allowed us to extend our results to higher values of $\kappa$.

% ********** Section **********
\section{Spot chimeras with localized coupling} \label{app_scwlc}
The spot chimera solutions for $A(\phi)$, $B(\phi)$, and $\Delta$ derived for $\kappa \ll 1$ can also be continued for larger values of $\kappa$.  However, unlike the spiral solutions, these solutions do not resemble sine functions, and as a result, they cannot be accurately represented using a truncated Fourier sine series.  Instead, a cosine series or full Fourier series may be appropriate. 

\begin{figure}
  \center
  \includegraphics[width=0.45\textwidth, trim = 0cm 0cm 0cm 0cm,clip=true]{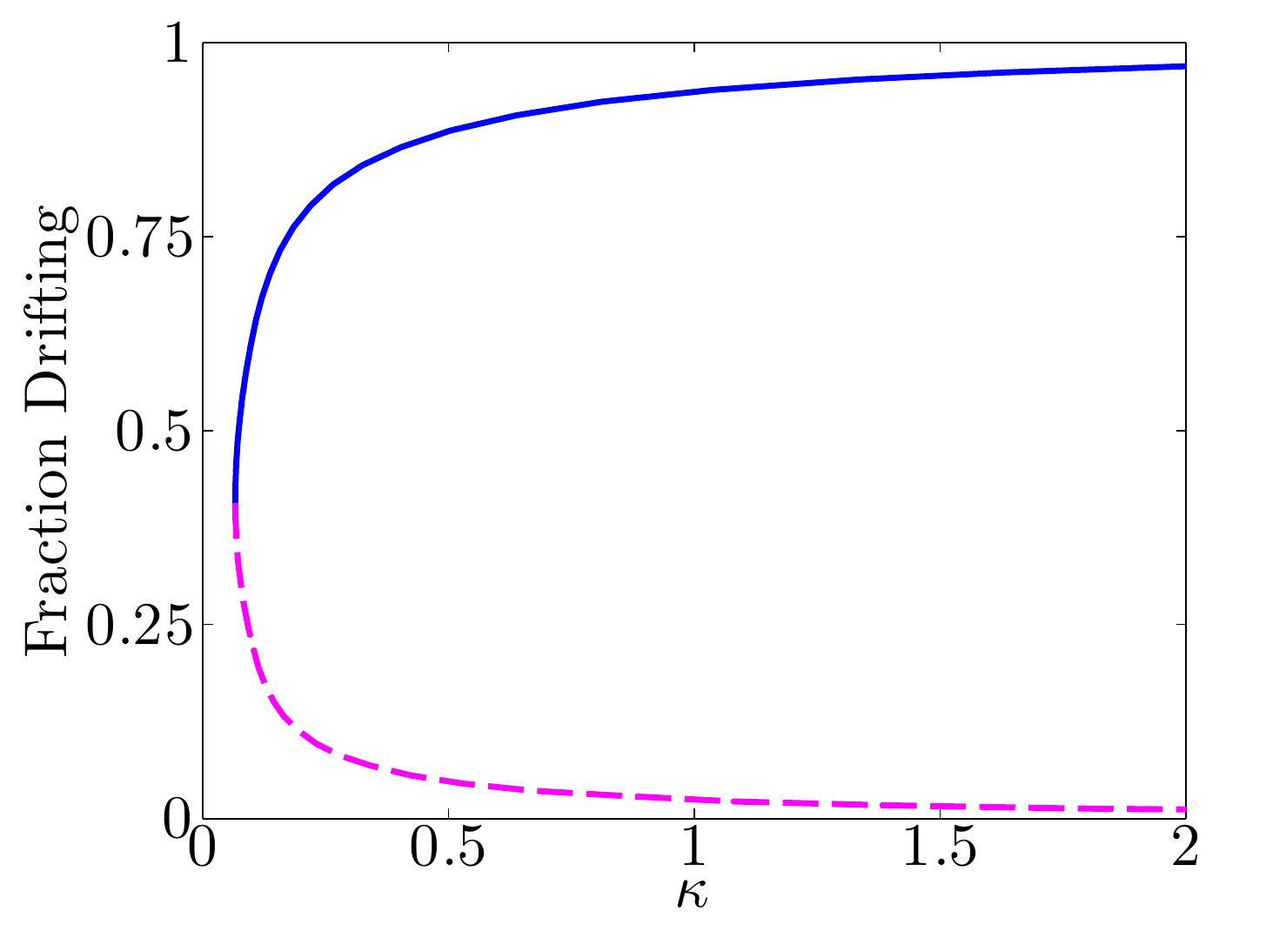}
  \caption{(Color online) Continuation of spot chimeras in $\kappa$. The blue (solid) curve is the continuation of a stable chimera and the magenta (dashed) curve is the continuation of an unstable chimera satisfying $\beta=0.01$.} 
  \label{fig:cont_spot}
% exact beta values are 0.010478 (stable),0.010912 (unstable)
\end{figure}

Numerical continuation results from MatCont (see Fig.~\ref{fig:cont_spot}) suggest that solutions exist for higher values of $\kappa$ that are qualitatively similar to the solutions with $\kappa \ll 1$. As the asymptotic analysis in section \ref{sec:global} indicates, the critical value of $\beta$ corresponding to the saddle node bifurcation grows with $\kappa$.  

In our numerical exploration of the stability of spiral chimeras, while integrating Eq.~\eqref{kuramoto} we observed that some of the unstable chimeras with $\alpha$ near $\pi/2$ and $\kappa>1$ evolved into spot chimeras. This suggests that spot chimeras remain stable for larger values of $\kappa$. As a result, we believe spot chimeras with localized coupling warrant further exploration. 

% ********** Section **********
\section{Generating initial conditions} \label{app_gic}
In many systems, the basin of attraction for a chimera state is small compared with the basins of attraction of the uniform coherent and incoherent states. Since chimera states can only be observed in simulations when the initial condition is inside these basins, completely random initial conditions are unlikely to converge to a chimera.

It is also difficult to determine the structure of the basins of attraction of equilibrium points in high dimensional systems. So, the most effective method for observing a chimera states in simulation is to select an initial condition very close to the chimera.  Such an initial condition can be found given a solution for the order parameter $R(\rvec)e^{i\Psi(\rvec)}$ and the natural frequency in the rotating frame $\Delta=\omega-\Omega$ by using the method outlined below.  

After making the transformation $\psi\rightarrow\psi+\Omega t$ to shift into a rotating frame of reference, Eq.~\eqref{kuramoto} can be written as 
\begin{equation*}
  \label{kuramoto_opversion}
  \frac{\partial\psi(\rvec)}{\partial t} = \Delta -R(\rvec) \sin (\psi(\rvec)-\Psi(\rvec)+\alpha).
\end{equation*}
This equation admits two types of stationary solutions. Wherever $R\geq\left|\Delta\right|$, oscillators have a stationary phase  $\psi^{*}$ satisfying
\begin{equation}
\label{stationaryphase}
\Delta -R\sin (\psi^{*}-\Psi+\alpha)=0.
\end{equation}
Wherever $R<\left|\Delta\right|$ oscillators cannot become phase-locked. However, if the phase $\psi$ at each point $\rvec$ is interpreted as a probability distribution $f(\psi)$, then there are stationary phase distributions satisfying the continuity equation 
\begin{equation*}
  \label{continuity}
  \frac{\partial f}{\partial t} +\frac{\partial}{\partial \psi}\left(f v\right)=0.
\end{equation*}
where $v=\Delta -R\sin (\psi-\Psi+\alpha)$ represents the phase velocity.  It is straightforward to check that the following distribution is stationary:
\begin{equation}
\label{stationarydist}
f(\psi)=\frac{\sqrt{\Delta^2-R^2}}{2\pi\left|\Delta-R\sin (\psi-\Psi+\alpha)\right|}.
\end{equation}
Therefore, given $R$, $\Psi$ and $\Delta$ at the position of each oscillator $\rvec$, an appropriate initial phase $\psi$ can be computed as follows: 
 
\begin{itemize}
\item[] \textit{Case 1.} If $R\geq\left|\Delta\right|$ set $\psi=\psi^{*}$ by solving Eq.~\eqref{stationaryphase}.
\item[] \textit{Case 2.} If $R<\left|\Delta\right|$ choose $\psi$ randomly using the probability distribution in Eq.~\eqref{stationarydist}.
\subitem (a) Compute the cumulative distribution $F(\psi)=\int_{-\pi}^{\psi}f(x)dx$.
\subitem (b) Choose $X$ to be a uniformly distributed random number between 0 and 1.
\subitem (c) Set $\psi=F^{-1}(X)$. 
\end{itemize}

\begin{figure}[b]
\includegraphics[width=0.45\textwidth, trim = 0cm 0cm 0cm 0cm, clip=true]{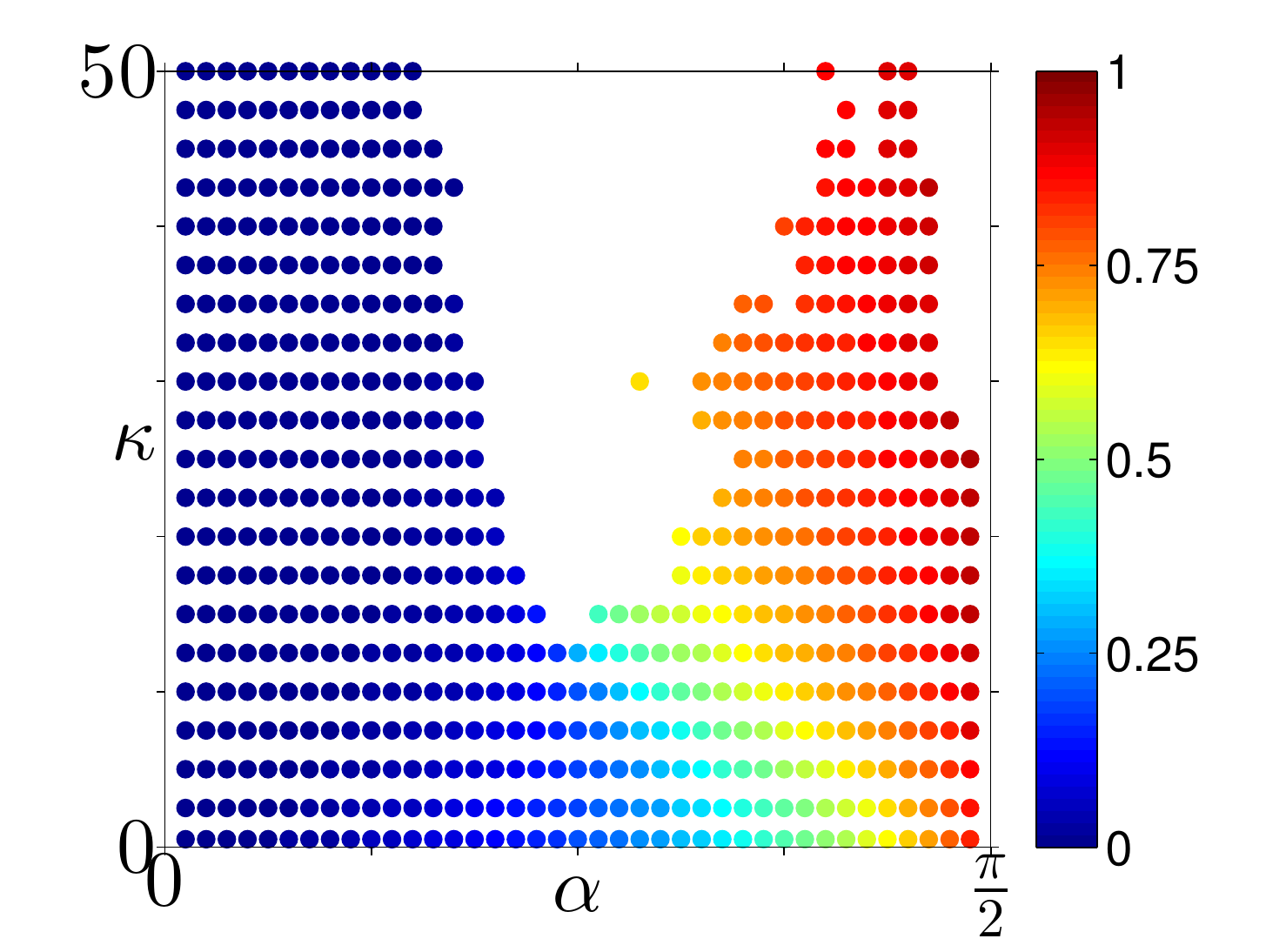}
\caption{(Color online) Fraction drifting for spiral chimera states. Dots represent points where spiral chimera states satisfying  Eq.~\eqref{new_op} exist. These points are identical to Fig.~\ref{fig:spiral_stability}. The color indicates the fraction drifting for each solution.  }
\label{fig:fracs}
\end{figure}

\begin{figure*}[t]
\includegraphics[width=0.75\textwidth, trim = 0cm 0cm 0cm 0cm, clip=true]{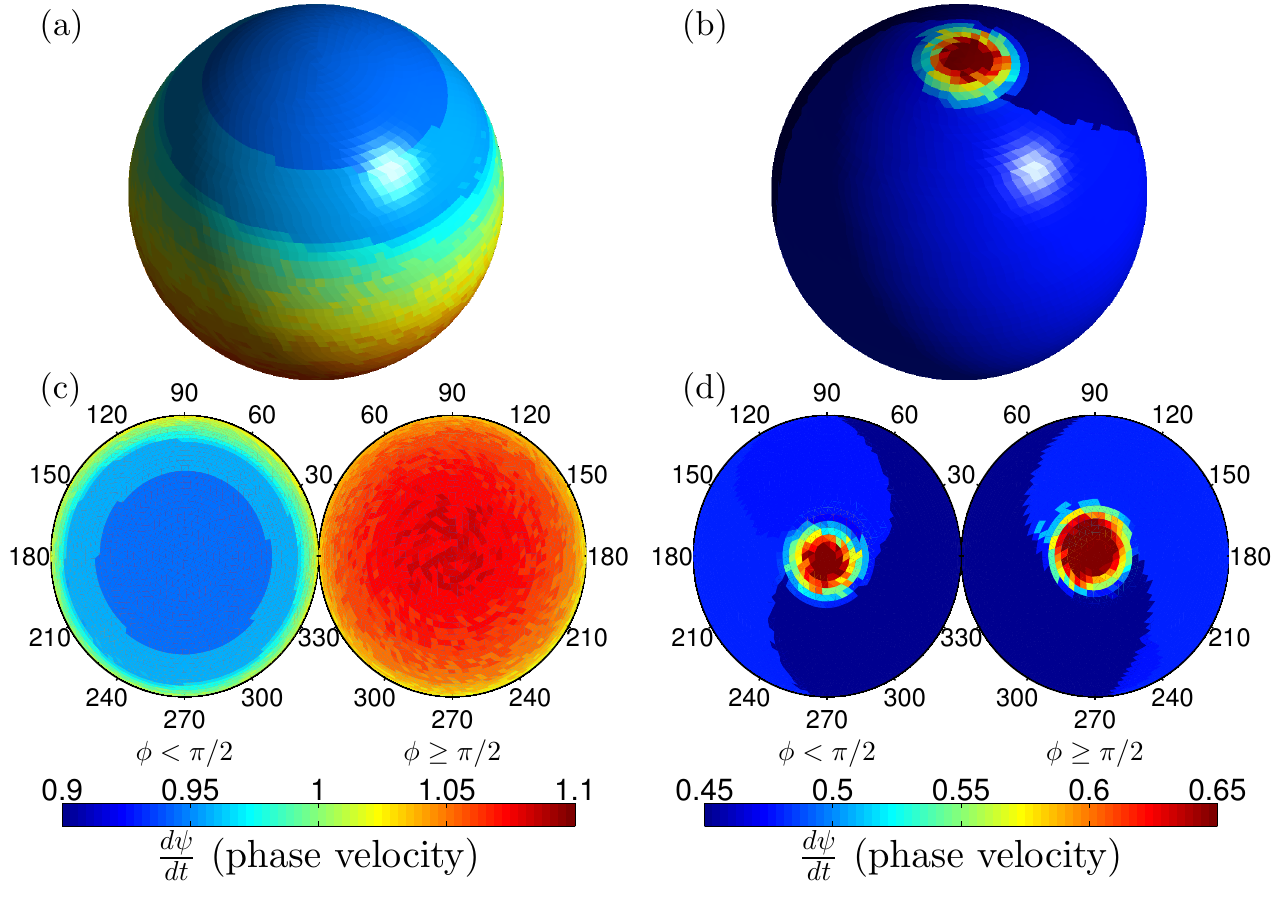}
\caption{(Color online) Phase velocities for spot and spiral chimera states. {Panels display} the phase velocities of 5000 oscillators (indicated by the color) corresponding to {a stable spot chimera} with $\Delta= 0.946$, $\alpha = 1.554$, and $\kappa = 0.25$ [panels (a) and (c)] and {a stable spiral chimera} with $\Delta = 0.478$, $\alpha= 0.589$, and $\kappa = 15$ [panels (b) and (d)]. Panels (a) and (b) display the sphere in three-dimensions, while panels (c) and (d) display two-dimensional projections of a sphere from above and below. The phase velocities are averaged over 200 units of time in a rotating frame in which locked oscillators have phase velocity $\Delta$. }
\label{fig:vels}
\end{figure*}

% ********** Section **********
\section{Numerical stability} \label{app_ns}

To complement our numerical results, we would have liked to perform a rigorous stability analysis on our system, as Omel'chenko was able to do for a ring of coupled oscillators \cite{Omelchenko2013}.  Unfortunately the nonlinear eigenvalue problem in our system results in a complex nonlinear integral equation with a nonseparable kernel which we were unable to solve analytically. 

There are various limitations to using numerical integration to ascertain information about stability in this system.  First of all, numerical integration itself introduces error.  To address this, we used MATLAB's built in adaptive Runge-Kutta method ODE45 for integration and verified that the results were consistent with those obtained using other ODE solvers. To accelerate computations, large matrix operations were carried out on a NVIDIA GeForce GTX 570 GPU with 480 cores using the Parallel Computing Toolbox, a MATLAB implementation of NVIDIA's CUDA platform.

Second, choosing uniformly distributed points on the surface of a sphere is a non-trivial problem.  There are various heuristic methods for generating points that are approximately uniformly distributed. We used the method described in Ref.~\onlinecite{Thomsen2007} which is a modification of the generalized spiral points method proposed by Rakhmanov \etal \cite{Rakhmanov1994}. By selecting evenly spaced points along a spiral from the north pole to the south, one obtains nearly ``uniformly'' distributed points. The slight nonuniformity means that some oscillators may be weighted slightly more heavily than others in the network. 
  
Third, the lifetime of chimera states can depend on the number of grid points. Previous work with spot chimeras has suggested that they are stable states with an infinite number of oscillators and long-lived transients with a finite number \cite{Omelchenko2011_2}. The lifetime of these metastable chimera states grows with the number of oscillators, but so does the computation time. For the figures displayed in this paper, we chose to use 5000 oscillators as a compromise allowing for full exploration of parameter space in a reasonable amount of time.  { Our results were robust to variations in the number of oscillators (we also tried 2500, 8000, and 10000 for selected cases of interest).} 
 However, there is no guarantee that the stability with a finite number of oscillators will agree with that when $N \to \infty$.

Note that even when the number of grid points is large, for some parameter values (e.g., small $\alpha$ for spiral chimeras) the number of points within the incoherent region may still be small (see Fig.~\ref{fig:fracs}), leading to inaccurate numerical assessment of stability (this is the origin of the yellow region in Fig.~\ref{fig:spiral_stability}). Furthermore, the effective coupling length given by \eqref{vm_kernel} is proportional to $\kappa^{-1}$, so the number of grid points must grow with $\kappa$ if a minimal number are to be included within the ``coupling zone'' {where coupling strength is significant}.

Finally, numerical integration cannot truly determine stability.  We integrated Eq.~\eqref{kuramoto} for 5000 units of time.  This termination criteria is somewhat arbitrary and could lead to unstable but long-lived transient states being classified as stable. In our analysis, we found that the boundaries of the ``stable'' region did change slightly depending on the termination criteria. However, we also tested a subset of points in the domain and verified that they appeared stable after 10000 units of time.  Both 5000 and 10000 time units are orders of magnitude longer than the transient lifetime of a typical chimera state that we identify as numerically unstable.

Figure \ref{fig:vels} shows the typical pattern of phase velocities distributed on the sphere for both stable spot and stable spiral chimera states. The final pattern of phase velocities was used in conjunction with the final pattern of phases to distinguish between stable and unstable chimera states.

%\bibliographystyle{naturemag}	% (uses file "plain.bst")
%\bibliography{chimera_sphere}		% expects file "myrefs.bib"

\begin{thebibliography}{10}
\bibitem{footnote}\bibinfo{note}{Another type of closed two-dimensional surface that may be relevant to physical systems is a torus which was studied in Ref.~\cite{Panaggio2013}.}
\end{thebibliography}

\end{document}